\documentclass[sigconf,authorversion=true]{acmart}
\usepackage{booktabs}
\usepackage{float}
\usepackage{listings}

\copyrightyear{2017} 
\acmYear{2017} 
\setcopyright{acmcopyright}
\acmConference{IML '17}{October 17-18, 2017}{Liverpool, United Kingdom}\acmPrice{15.00}\acmDOI{10.1145/3109761.3109765}
\acmISBN{978-1-4503-5243-7/17/10}

\begin{document}

\title{
  An Email Attachment is Worth a Thousand Words, or Is It?
}

\author{Gregory Tsipenyuk}
\email{gregory.tsipenyuk@cl.cam.ac.uk}
\affiliation{%
  \institution{University of Cambridge Computer Laboratory}
  \streetaddress{William Gates Building, 15 JJ Thompson Avenue}
  \city{Cambridge}
  \country{United Kingdom}
  \postcode{CB3 0FD}
}
\author{Jon Crowcroft}
\email{jon.crowcroft@cl.cam.ac.uk}
\affiliation{%
  \institution{University of Cambridge Computer Laboratory}
  \streetaddress{William Gates Building, 15 JJ Thompson Avenue}
  \city{Cambridge}
  \postcode{CB3 0FD}
  \country{United Kingdom}
}

\begin{abstract}
There is an extensive body of research on Social Network Analysis (SNA) based on the email arhive. 
The network used in the analysis is generally extracted either by capturing the email communication in From, To, Cc and Bcc email header fields or by the entities contained in the email message. 
In the latter case, the entities could be, for instance, the bag of words, url's, names, phones, etc. 
It could also include the textual content of attachments, for instance Microsoft Word documents, excel spreadsheets, or Adobe pdfs. 
The nodes in this network represent users and entities. The edges represent communication between users and relations to the entities. 
We suggest taking a different approach to the network extraction and use attachments shared between users as the edges.
The motivation for this is two-fold. 
First, attachments represent the ``intimacy'' manifestation of the relation's strength. 
Second, the statistical analysis of private email archives that we collected and Enron email corpus shows that the attachments contribute in average around 80-90\% to the archive's disk-space usage, which means that most of the data is presently ignored in the SNA of email archives. 
Consequently, we hypothesize that this approach might provide more insight into the social structure of the email archive.
We extract the communication and shared attachments networks from Enron email corpus. We further analyze degree, betweenness, closeness, and eigenvector centrality measures in both networks and review the differences and what can be learned from them.
We use nearest neighbor algorithm to generate similarity groups for five Enron employees. The groups are consistent with Enron's organizational chart, which validates our approach. 
\end{abstract}


\keywords{Social Network Analysis, email, email attachment}

\maketitle

\section{Introduction}
\label{s:introduction}

There is an extensive body of research on Social Network Analysis (SNA) based on the email communication. 
The subject of the research can cover multiple topics such as relation discovery~\cite{Shetty2005}, software project activity~\cite{Bird2006}, group inference~\cite{Yelupula2008}, hierarchy detection~\cite{Duczynski2015}, crisis analysis and prediction~\cite{Diesner2006}, topic and role discovery~\cite{McCallum2007}, text analysis of social values~\cite{Zhou2010},  fraud detection~\cite{Tang2010}, information extraction and search~\cite{Laclavik2012}, classification~\cite{Wang2011}, and SPAM detection~\cite{Lam2007}. 
Typically, the analyzed network reflects either communication between users or a relationship between the email and the information found in the email's header and the body. 
In the former, the nodes represent users found in the From, To, Cc, and Bcc email header fields, and edges, either directed or undirected, represent relationship between the sender (From) and the recipient (To, Cc, Bcc), with the weight reflecting the frequency of communication between the two, for instance as in~\cite{Diesner2006}. 
In the latter, the nodes represent the email and entities extracted from the email like people, phone number, email addresses, etc., and the edges represent co-occurrence of named entities within the email or its parts, including converted to text attachments, for instance as in~\cite{Laclavik2012}.

\begin{table*}
	\centering
	\caption{Attachment's statistics}
	\begin{tabular}{|l|l|l|}
	\hline
		                                    & Enron   & Friends \& Family \\ \hline
		Avg. Size of Attachments in Mailbox & 91.26\% & 81.93\% \\ \hline
		Avg. Messages w/Attachments in Mailbox        & 24.49\% & 14.26\% \\ \hline
		Type of Attachments in all Mailboxes & Docs: 94.58\% & Docs: 17.02\% \\
		                                     & Multimedia: 4.21\% & Multimedia: 79.2\% \\
		                                     & Other: 1.21\%      & Other: 3.78\% \\ \hline
	\end{tabular}
\label{t:attachstats}
\end{table*}

We analyzed private email archives of friends and family and the Enron email corpus. The statistics is shown in Table~\ref{t:attachstats}.
While the average number of messages with attachments is only 14.26\%(Private) and 24.49\%(Enron), the average disk space that they take is high 81.93\%(Private) and 91.26\%(Enron). Another observation is that 94.58\% of attachments in the Enron's corpus are documents and 79.2\% of attachments in the Private archive are Multimedia (Audio,Video,Image).   
Therefore, generally most of the data is discarded in SNA based on the communication network. 
While multidimensional network based on entities extracts textual content of attachments, it will ignore images, which may represent substantial part of the email message. 
Besides quantitative contribution to SNA, attachments may have a qualitative property as well by representing the ``intimacy'' manifestation of the relationship strength. 
Granovetter defined the strength of a tie in his seminal paper ``The strength of weak ties'' as ``The strength of a tie is a (probably linear) combination of the amount of time, the emotional intensity, the intimacy (mutual confiding), and the reciprocal services which characterize the tie''~\cite{Granovetter1973}. In ~\cite{Gilbert2009} authors show that ``intimacy dimension'' makes the largest contribution of 32.8\% to the tie strength prediction model based on social media. 
Indeed, it is plausible to assume that sharing, for instance, a picture of an interesting event or an important document might indicate a higher level of trust and close relationship between two people.

We are proposing to extract the social network based on email attachments shared between user accounts, constructing one-mode projection of bipartite graph as in ~\cite{Wang2012}. 
We view the attachment as a ``virtual event'' attended by users sharing the attachment. 
This is a variation of similar sentiment described in ~\cite{Gupte2012}: ``Each email is an event and all the people copied on that email \textendash i.e., the sender (From) and the receivers (To, Cc, Bcc) \textendash are included in that event.''. 
Indeed, if Alice sends pictures of a New Year party to Bob then Bob ``virtually'' shares some, possibly the most exciting, party's experience with Alice. Likewise, if Bob sends some important document to Alice then Alice collaborates on the document with Bob via the ``virtual'' meeting. 
We can then extract the social network with nodes representing users and edges representing the attachments shared between the users. 
The way we are inferring the sharing of attachments is via Secure Hash Algorithm 1 (SHA1)~\cite{Easlake2011} of the attachment's content. 
This can be accomplished by parsing each email message of each user's email archive into its MIME parts~\cite{Freed1996a}, aggregating SHA1 of attachments by the user, and finding common attachments between users. 
But not every attachment is indicative of a quality social relationship between users. For instance, some attachments could be a part of the bulk-email, making every user connected. Other cases include common logos, signatures, or Internet trends. Indeed, an intra-company's email with the company's logo attached will make all users connected. A methodology should be used to filter out not meaningful for the analysis attachments.
We hypothesize that the social network extracted from shared attachments will provide more insight into the relationships between people.  

The contribution of this paper is five-fold. 
First, we are proposing to extract the Social Network via shared email attachments. 
Second, we are building the email dataset based on Enron email corpus with attachments. 
Third, based on empirical evidence we are proposing a methodology of filtering out not meaningful attachments. 
Fourth, we extract the communication and shared attachments networks from Enron email corpus, analyze degree, betweenness, closeness, and eigenvector centrality measures in both networks and review their differences. Fifth, we apply nearest neighbor algorithm on the dataset of employees with the corresponding list of attachment's SHA1 in order to predict similarity lists for some employees.

The rest of the paper is organized as follows. 
Section \ref{s:relatedwork} reviews the related work. 
Section \ref{s:arcitecture} describes the dataset building. 
Section \ref{s:netextract} talks about network extraction. 
Section \ref{s:netanalysis} presents evaluation data and analysis. 
Finally, section \ref{s:conclusion} concludes.

\section{Related Work}
\label{s:relatedwork}

In ~\cite{Laclavik2011} ~\cite{Laclavik2012} authors construct multipartite graph from the email. 
Each email has its own node with connection to entities extracted from the email such as people, email addresses, phone numbers, dates, etc. 
Unique entities appear only once in the graph but are connected to each email where they appear. 
Edges are links between entities representing co-occurrence in the same email part, paragraph, sentence or a composite named entity. 
Attachments within the email are converted, where possible, to the textual representation. 
This excludes multimedia attachments like image, video, or audio, which could represent most attachments in private emails. 
In ~\cite{Guillaume2006} authors suggest that interactions of users and information in real world complex networks like Internet, Web, movie actors, co-authors, word's co-occurrence, and protein could be modeled with bipartite graph. In this graph users and information are represented by two disjoint vertex set and edges represent relationship between users and information.
In ~\cite{Wang2012} authors use one-mode projection of bipartite graph to capture the similarity of information and shared interests of users. 
They apply this methodology to on-line news aggregation site to analyze user's similarity in voting on news stories.

\section{Dataset}
\label{s:arcitecture}

In our analysis we used Shetty\&Adibi\cite{Shetty2004} dataset linked to Electronic Discovery Reference Model (EDRM)\footnote{\url{http://www.edrm.net}} Version Two Enron corpus dataset. 
The Enron email corpus was released by the Federal Energy Regulatory Commission during the investigation into Enron's collapse. 
William Cohen from CMU prepared the dataset and published it for researches\footnote{http://www-2.cs.cmu.edu/~enron}. 
The data set contains 517,424\footnote{This is 7 less than the originally released dataset of 517,431 emails. 7 emails were removed by William Cohen for privacy reasons.} emails from 151 users with 242,944 unique content's SHA1. 
Shetty\&Adibi removed duplicate messages from the dataset and fixed some email address discrepancies. 
Their dataset contains 252,759 email messages with 212,326 unique content's SHA1. 
To verify that Shetty\&Adibi is a subset of Cohen's dataset we first checked that Message-ID set from Cohen's dataset is a superset of Shetty\&Adibi's dataset. 
Second, we collected a Simple Random Sample (SRS) of 384 (per~\cite{Bertlett2001}) emails from Shetty\&Adibi's dataset and verified via content's SHA1 that emails match those in Cohen's dataset. 
As part of the verification we had to make some fixes in Shetty\&Adibi's email's body: 1) 14 messages had X-FileName header included in the content; 2) 1 message had its body truncated. 
Neither Cohen's or Shetty\&Adibi's dataset contains attachments. 
EDRM released two versions of the Enron's dataset with attachments. 
Version One (EDRMV1) contains 697,079 emails with 155,431 unique content's SHA1 for 130 users. 
Version Two contains 1,234,387 emails with 242,800 unique content's SHA1 for 151 users. 
EDRMV1 does not provide continuity and applicability of our analysis to the previous body of research because the data is missing for 21 employees. 
We therefore decided to use EDRMV2 in our analysis. 
However, we discovered that email addresses in From, To, Cc, and Bcc header fields in EDRMV2 don't conform to canonical email address format of \textit{user@domain} and are most likely taken from the original Enron email corpus. 
Here are some instances of Phillip Allen's email address:

\hfill \break
\noindent
\textit{
"ALLEN  PHILLIP K" <pallen@enron.com> \\
"phillip.k.allen" <phillip.k.allen@enron.com> \\
<Allen>,"Phillip" \\
<Allen>,"Phillip K." \\
<Phillip.Allen@enron.com> \\
<Allen>,"Phillip K." </O=ENRON/OU=NA/CN=RECIPIENTS/CN=Pallen> \\
Phillip K Allen <Phillip.K.Allen@enron.com> \\
Allen, Phillip K. \\
Phillip K Allen \\
Phillip,K,Allen \\
}

Those formats are partially consistent with what was found in ~\cite{Zhou2007}.
However, the problem is compounded by the email lists having comma separated addresses in mixed format.
Here are some examples:

\hfill \break
\noindent
\textit{
Brad,Alford,Phillip,K,Allen,anderson,Bob \\
Kristin Albrecht, Phillip K Allen, Hunter S Shively \\
Frolov, Yevgeny </O=ENRON/OU=NA/CN=RECIPIENTS/CN=Yfrolov>, Allen, Phillip K.\\
}

The parsing of these lists is not trivial and error prone. 
We therefore decided to take EDRMV2 dataset and replace From, To, Cc, and Bcc header fields with the ones from Shetty\&Adibi's dataset. 
We are linking emails from two datasets via employee, folder, subject, date, and content's SHA1 key. 
We cannot link on Message-ID because EDRMV2 has its own generated Message-ID. 
Our key uniquely identifies 252,722 emails in Shetty\&Adibi and 754,906 emails in EDRMV2. 
However, Date header field in Shetty\&Adibi's dataset doesn't include Time Zone information and we noticed that EDRMV2 has, in some instances, the Date changed to a different Time Zone, so the date without the Time Zone will not match in otherwise identical emails in Shetty\&Adibi and EDRMV2 datasets. 
Our solution is to link Cohen's dataset to EDRMV2's dataset with dates converted to GMT and filter each email by the Message-ID from Shetty\&Adibi's dataset. 
For simplicity, we will refer in the text below as simply Shetty\&Adibi dataset rather than Cohen's filtered by Shetty\&Adibi.

We had to make the following data preparation in order to achieve the best linking of Shetty\&Adibi to EDRMV2 dataset:
\begin{itemize}
\item
We calculate SHA1 in both datasets by extracting the email's content, removing all new lines and spaces, mapping all quotable-printable\footnote{\url{https://en.wikipedia.org/wiki/Quoted-printable}} characters to a '?' and then replacing all multiple occurrences of '?' to a single '?'. 
This has to be done because EDRMV2 has the content reformatted in many instances. 
To get the SHA1 for EDRMV2 consistent with Shetty\&Adibi SHA1 we remove ``boundary'' lines from EDRMV2 content and remove all content following and including a copyright notice inserted by EDRM.
\item
Some content in EDRMV2 is truncated. 
We matched 1,990 emails from Cohen's dataset by truncating emails to the same length as in EDRMV2 dataset. 
We set the minimum length of the content in those cases to 100 bytes to provide for sufficient entropy in the text.
\item
We had to make changes to match dates between two datasets. 
In EDRMV2 when the email's folder is ``schedule\_crawler'', the time is 4 hours behind of corresponding Shetty\&Adibi time. 
In addition, we found the following pattern where EDRMV2 time is behind Shetty\&Adibi time by the number of hours specified in the zone difference with UTC or by 2, 3, 4, 10 and 12 hours. 1,225 linked emails had the date fixed this way.
\item
For 8,627 messages we had to ``downgrade'' the key to subject, date, and SHA1.
\end{itemize}

Overall 249,353 out of 252,754 emails were linked, which represents 98.6\% of emails from Shetty\&Adibi's dataset. 3,401 emails were not linked. 
The missing emails are distributed over 125 employees. 
Further analysis shows that top 10 from that list account for 2,312 of missing emails or about 68\%. 
Top 2 out of 10, Richard Sanders and Jeff Dasovich account for 51\% and 17\% respectively. 
Judging by the SHA1 analysis, the missing employees from Shetty\&Adibi's dataset don't have corresponding emails in EDRMV2 dataset. 
Based on SHA1 analysis up to 1,630 missing emails can be recovered from EDRMV1. 
Due to time limitations we left this work for the future research.

\section{Network Extraction} 
\label{s:netextract}

We construct the graph to analyze the Enron social network in two ways. 
First is via communication between users, where nodes represent the email address or the user listed in From, To, Cc, and Bcc email header fields. 
Edges represent the relationship between the sender (From) and recipients (To, Cc, Bcc). 
Edges are undirected and weighted by the frequency of communication. 
\begin{figure}[h]
	\centering \includegraphics[width=\columnwidth]{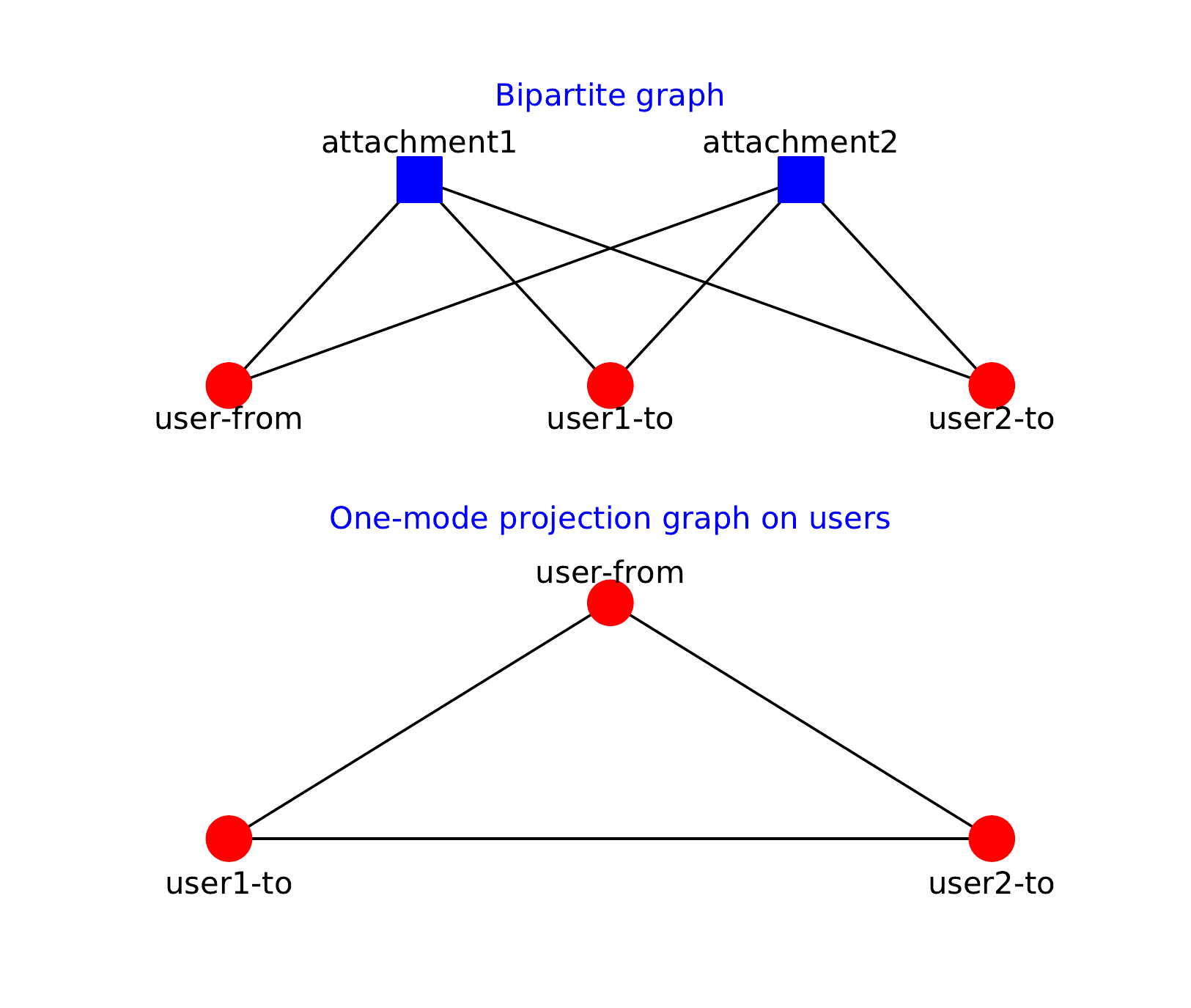}
	\caption{\label{f:bipartite}Bipartite graph with attachments and users vertices}
\end{figure}
Second is via shared attachments by constructing one-mode projection graph on users of bipartite graph as demonstrated on Figure~\ref{f:bipartite}. The top figure shows bipartite graph extracted from an email sent from \textit{user-from} to two users \textit{user1-to} and \textit{user2-to}. The email contains two attachments \textit{attachment1} and \textit{attachment2}. The bottom figure shows corresponding one-mode projection graph on users, where nodes represent users. And edges represent attachment sharing between users. The edges are undirected and weighted by the number of shared attachments. The edges are undirected because not in all cases the direction is meaningful. Consider a user \textit{U} sending an email with an attachment \textit{A} to users \textit{U1} and \textit{U2}. In the extracted one mode projection graph on users, the edges \textit{\{U,U1\}, \{U,U2\}} can have the direction but the edge \textit{\{U1,U2\}} can not. Moreover, it is possible that a user \textit{U1} sends an email with an attachment \textit{A} to a user \textit{U2} and another user \textit{U3} independently sends an email with the same attachment \textit{A} to a user \textit{U4}. In this case edges \textit{\{U1,U2\},\{U3,U4\}} can have the direction and edges \textit{\{U1,U4\},\{U2,U4\},\{U1,U3\},\{U3,U2\}} can not have the direction. 

To extract the graph from an archive,
for each user \textit{U} archive \textit{A} we extract email messages which contain attachments. 
For each attachment in the extracted email a global dictionary is maintained keyed by 1) the attachment; 2) user \textit{U} and every user \textit{V} in the email's From, To, Cc, and Bcc header fields distinct from the user \textit{U}; 3) the email's Message-ID. 
Then for each attachment in the global dictionary we look at all unique user pairs and create nodes for these users, if nodes don't already exist, and the edge connecting these nodes. 
By aggregating sender and recipient users in the same pool we are not only capturing direct communication between users, for instance a user in From sends email to a user in To, but also capturing a friend of a friend relationship, i.e. all users in To, Cc, and Bcc who may not communicate directly but become connected via the shared attachments. 
In addition, by capturing a user \textit{U}, the owner of the archive \textit{A}, we may capture relationships of the user \textit{U} to users in From, To, Cc, and Bcc when none of these users is the user \textit{U}. 
We extract both networks for core 151 employees only. 

\begin{figure}[h]
	\centering \includegraphics[width=\columnwidth]{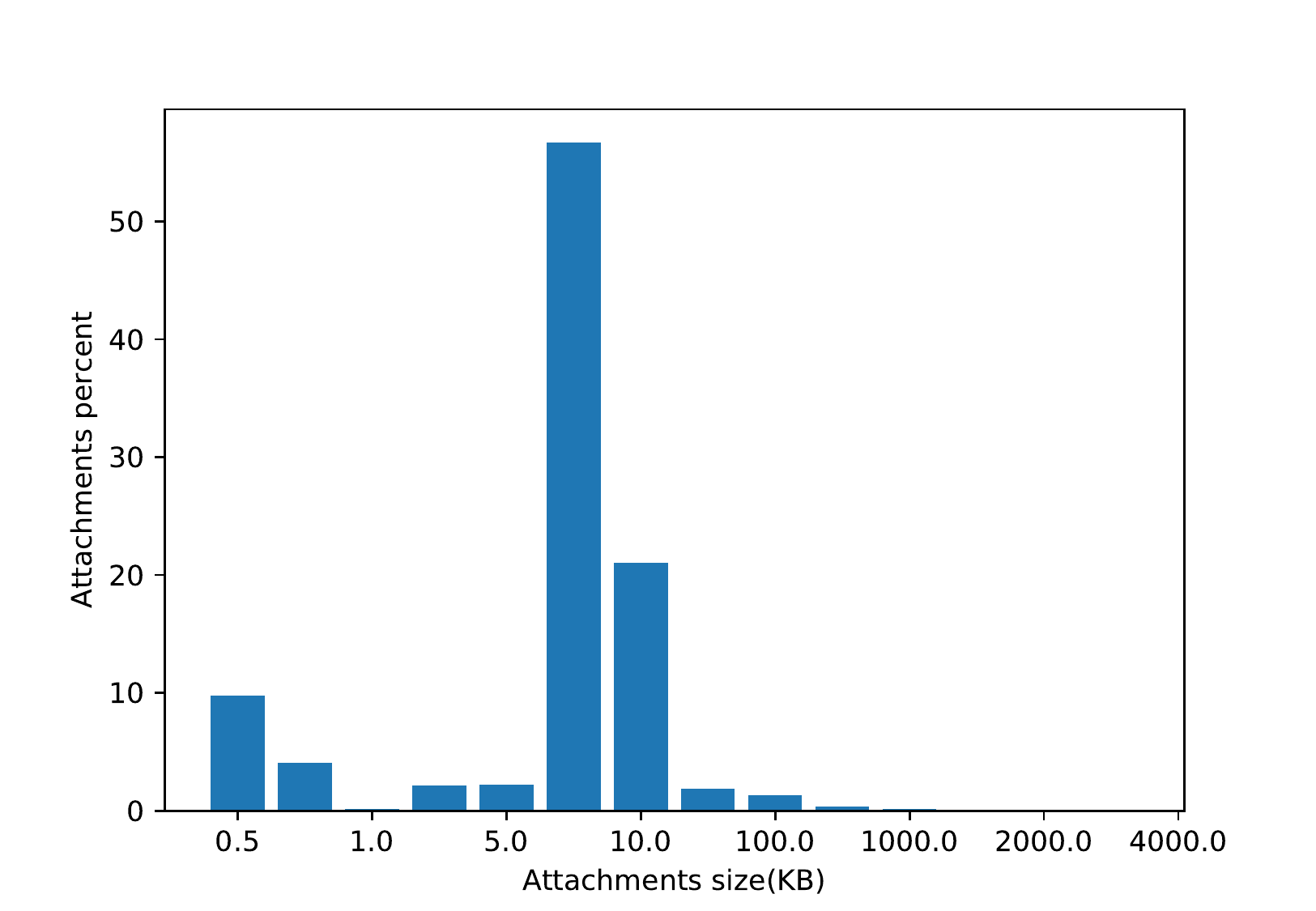}
	\caption{\label{f:size}Attachments size histogram.}
\end{figure}

An important part of extracting the Social Network is defining the strength of ties or the weight of edges. 
As noted in ~\cite{Choudhury2010}, the tie strength itself remains an ambiguous concept with multiple, possibly inconsistent definitions and that there is a non-trivial range of thresholds of 5-10 reciprocated emails per year which maximizes prediction of relevant task that depends on various network features. 
Some researchers simply count the frequency of communication as for instance in~\cite{Agarwal2012}. 
Others consider Cc communication less important than To and decrease it at an inverse square-root rate~\cite{Kaye2014}. 
And yet others have the email frequency threshold, which is typically set to 5 emails~\cite{Shetty2004}. 
How the tie strength could be defined when the Network is extracted from shared attachments? We don't think that the size of the attachment matters. 
Indeed, an event is shared via the attachment regardless of the image resolution or the document length.
Consequently, we suggest assigning the weight of 1 to the attachment. 
But should the weight be aggregated over all attachments in the email or should it be 1 regardless of the number of attachments? Either approach is sensible. 
Suppose Alice sends 10 New Year Party pictures to Bob. 
Alice and Bob share one event so the weight should be 1. 
But what if Alice sends 10 pictures from 10 different events in the same email? Then the weight should be 10. 
We are going to make an assumption that an email generally contains attachments related to one event and one shared attachment creates one tie between two users. 
All other attachments from the same unique email are ignored. 
But if an attachment from the email appears in another unique email then this attachment will be considered as another tie. 
We'll defer developing of a better methodology for the tie strength to future research.

\begin{figure}[h]
	\centering \includegraphics[width=\columnwidth]{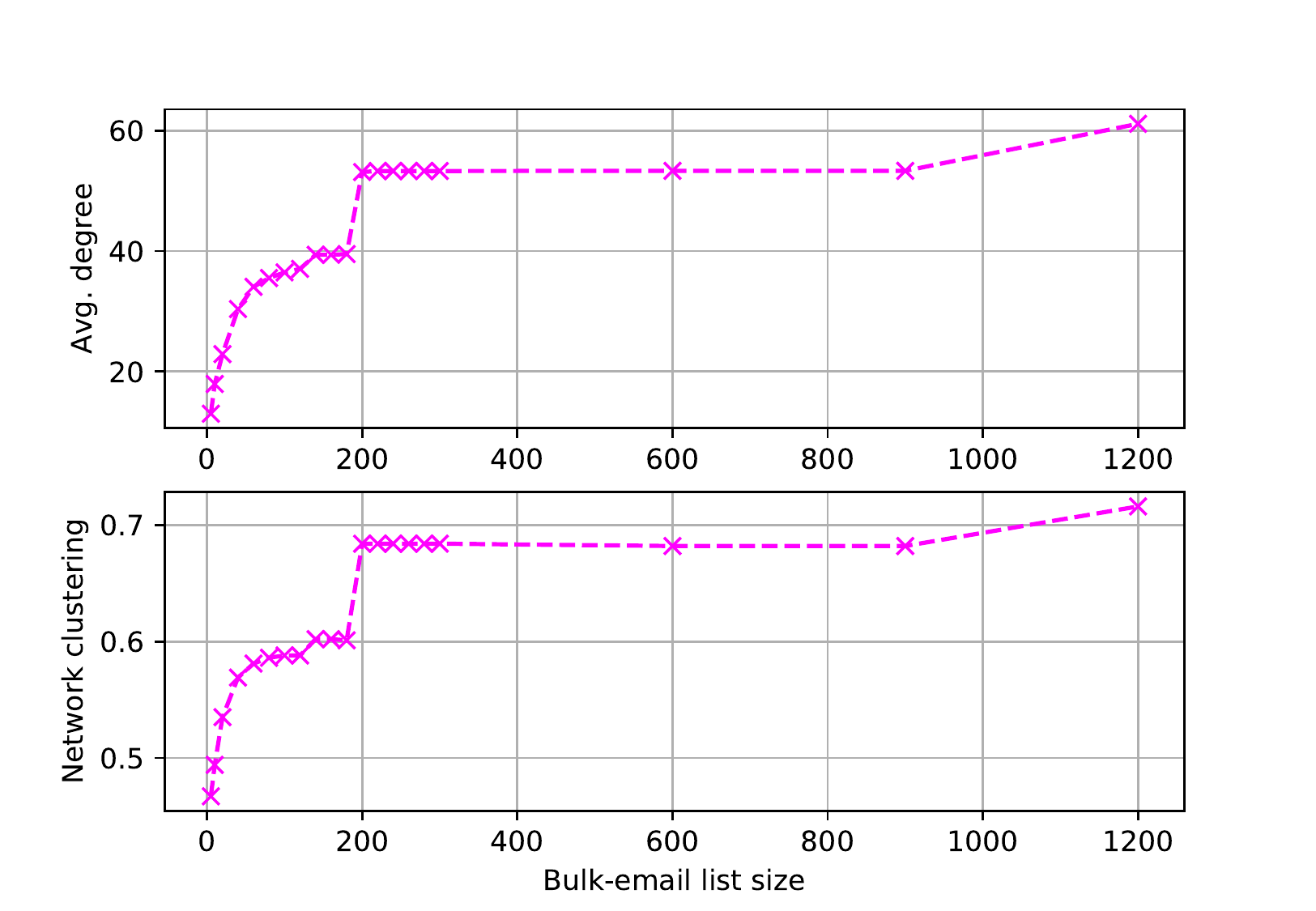}
	\caption{\label{f:bulk}Avg. degree and clustering depending on the filtered bulk email}
\end{figure}

\begin{figure}[h]
	\centering \includegraphics[width=\columnwidth]{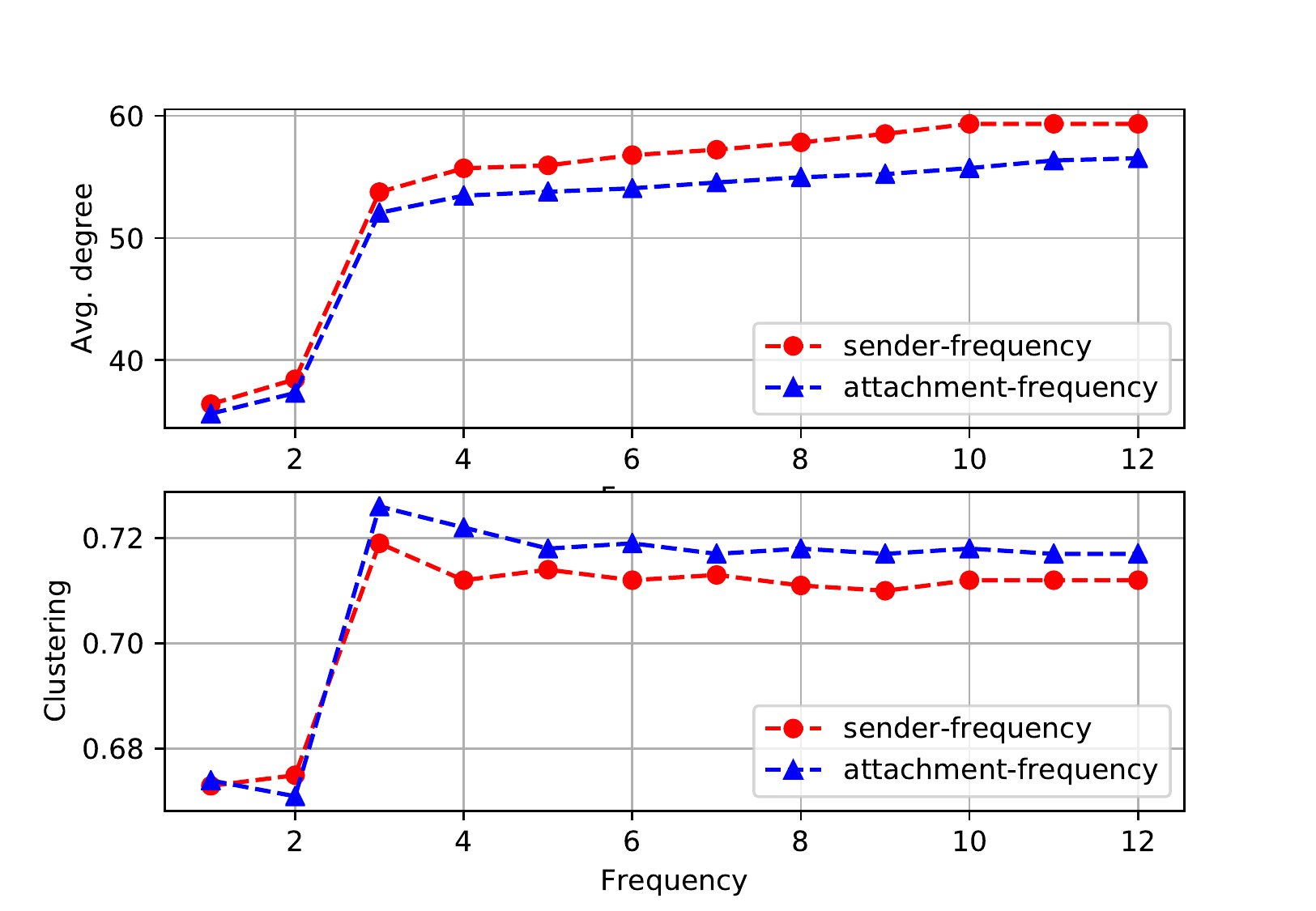}
	\caption{\label{f:frequency}Avg. degree and clustering depending on the filtered frequency of the attachment and the sender}
\end{figure}

Another issue to consider when extracting the network from shared attachments is the value of the attachment for our analysis. 
For instance, with present day proliferation of e-commerce and growth of the machine-generated emails, it is possible that many users have the same e-site logos in their email messages, like Amazon logo. This doesn't make all users of Amazon  socially connected. 
Overall, this kind of attachment could be a company logo, an e-signature, a common document like a benefit
form, and an Internet trend or rumour where a trendy multimedia or a document are spread to many users in the network. 
We jointly call these attachments TRAM (TRend+spAM). 
How can we filter out TRAM attachments? 
We are suggesting the following approaches to filtering out non-useful attachments:
\begin{itemize}
\item
  A threshold on the attachment's size may filter out common logos and e-signatures. The thought is that these type of attachments generally has a small size. Figure~\ref{f:size} shows attachment's size histogram. There are two spikes at 0.1K and 0.5K. We reviewed SRS of attachments with the size less than 1K. We found that all attachments except for one were TRAM. Most of these attachments are artifacts of the EDRM building of the Enron dataset with attachments, not our linking processing. Those attachments are either documents with the text stating that the attachment's link was not found or documents that we cannot open. Others are e-business cards, logos, executable files, news and e-site urls. We ignore an attachment if its size is less or equal than 1K. The extracted network has average degree 59.6 and clustering 0.714 as opposed to unfiltered network with average degree 61.12 and clustering 0.716. Removing small size attachments has small effect on the network connectivity.
\item
  Bulk email can significantly effect Network's connectivity and clustering. Indeed, sending a broadcast message with the attachment of the company's quarterly earnings will make all employees share the same attachment. But it will not make all of them socially connected. Figure~\ref{f:bulk} shows average degree and clustering depending on the filtered bulk email list size. We can see two interesting points at about 200 and 900 list size. They are explained by 4 emails sent to 193 users and 2 and 3 emails sent to 947 and 948 users respectively. Clearly 900 can be put in the bulk email group. But it is not obvious what the low threshold should be. Dunbar in~\cite{Dunb} suggested social group sizes of 5,12,35,150,500, and 2,000 with 150 being a cognitive limit also known as Dunbar's number. Hill and Dunbar in~\cite{Hill2003} further categorized the groups as support cliques, sympathy groups, bands, and higher-level groupings (above 35). In~\cite{Dunbar1993}, Dunbar defined band group size as 30-50 individuals. Dunbar then suggested in~\cite{Dunbar2011} that there are top 50 (corresponding to the band group) with whom we keep up every month or so and then there are all others with whom we correspond in any meaningful way. The 200 email size falls into higher level grouping and can be categorized as the bulk email. We therefore are suggesting to set the band group size of 35 as the bulk email threshold. This threshold corresponds to the extracted network with average degree 29.4 and clustering 0.566. This is a change of 52\% and 21\% for average degree and clustering respectively.
\item
  Other properties that influence Network's connectivity and clustering are the frequency of the shared attachment in unique emails and the frequency of unique senders of the shared attachment. Consider the Amazon logo example we mentioned above - many users have the same logo attachment in their unique emails. Another example is an intra-company communication with email containing the company's logo. In this case many unique emails have the same attachment but also the same attachment is sent by many unique senders. A news story or a rumour-trend gone viral with the message being re-sent by multiple users is similar to the company's logo example. Figure~\ref{f:frequency} shows average degree and clustering depending on the filtered attachment and sender frequency. There is a point at the frequency 3 with a sharp change in the average degree from 35 to 53 and the clustering from 0.67 to 0.72. Number 3 is not coincidental for both the attachment and the sender frequency. In the former, the attachment in three unique messages with three unique senders or receivers creates the closed triad. And in the latter, the attachment with three unique senders always creates a closed triad. Consequently, we see higher clustering and degree when the frequency is greater than 3. Attachment's frequency has higher clustering because the set of attachments with frequency three or higher is a superset of senders frequency with three or higher. Indeed, the attachment with three unique senders will only occur in unique messages. We filter out attachments with frequencies higher than 2. The network extracted this way has average degree of 37.32 and clustering of 0.671.
\end{itemize}

We note that the above approach at best establishes the low boundary of described properties and may eliminate some common logo, signature, or trend attachments but doesn't ensure the quality of an attachment and also removes some valuable attachments in the process. It generally limits the cases of the attachment sharing to a single email sent to a list of at most 35 users or emails forwarded only once. 
A better solution might be to use methodology similar to SPAM detection based on Social Networks as in~\cite{Lam2007}, unsupervised Machine Learning, or information spread. We leave this research to future work.

\section{Network Analysis}
\label{s:netanalysis}

\subsection{Graph and Node Statistics}
\label{s:graph_node_stats}

\begin{table*}
\centering
\caption{General statistics for Communication and Attachments Networks.}
\begin{tabular}{|p{4cm}|p{4cm}|p{4cm}|p{4cm}|}
	\hline
	Statistics                 & Communication Network & Attachments Network, filtered & Attachments Network, unfiltered \\ \hline
	Clustering coefficient     & 0.545                 & 0.566                         & 0.716                           \\
	Connected components       & 1                     & 1                             & 1                               \\
	Network diameter           & 4                     & 5                             & 3                               \\
	Network radius             & 2                     & 3                             & 2                               \\
	Network centralization     & 0.448                 & 0.287                         & 0.407                           \\
	Characteristic path length & 2.025                 & 2.295                         & 1.62                            \\
	Avg. number of neighbors   & 24.12                 & 20.04                         & 61.12                           \\
	Number of nodes            & 150                   & 149                           & 150                             \\
	Number of edges            & 36,485                & 11,408                        & 33,780                          \\
	Network density            & 0.162                 & 0.135                         & 0.41                            \\
	Network heterogeneity      & 0.588                 & 0.567                         & 0.483                           \\ \hline
\end{tabular}
\label{t:generalstats}
\end{table*}

\begin{table*}
\centering
\caption{Top 10 employees for centrality measures in Communication Network}
\begin{tabular}{|l|l|l|l|}
	\hline
	Overall Rank & Name            & Title                      & Centrality Measures                \\ \hline
	1            & John Lavorato   & CEO, ENA                   & D(7),EV(1),B(2),C(1),T(1)          \\
	2            & Liz Taylor      & Assistant to President     & D(55),EV(2),B(1),C(2),T(2)         \\
	3            & Louise Kitchen  & President, EO              & D(9),EV(4),B(6),C(5),T(4)          \\
	4            & Sally Beck      & COO                        & D(48),EV(3),B(3),C(3),T(3)         \\
	5            & Kenneth Lay     & CEO                        & D(50),EV(5),B(5),C(4),T(5)         \\
	6            & Jeff Dasovich   & Dir Stat Gov Affairs       & D(1),EV(23),B(8),C(11),T(16)       \\
	7            & Phillip Allen   & Managing Director Trading & D(33),EV(6),B(15),C(6),T(6)        \\
	8            & Kevin Presto    & VP Trading, ENA East Power & D(30),EV(7),B(11),C(7),T(7)        \\
	9            & Mike Grigsby    & VP Trading, ENA Gas West   & D(11),EV(10),B(16),C(8),T(8)       \\
	10           & Scott Neal      & VP Trading, ENA Gas East   & D(38),EV(8),B(20),C(9),T(9)        \\
	11           & David Delainey  & CEO ENA\&EA                & D(17),EV(9),B(28),C(10),T(14)      \\
	12           & Tana Jones      & Senior Legal Specialist    & D(2),EV(26),B(29),C(22),T(22)      \\
	13           & James Steffes   & VP Government Affairs      & D(3),EV(21),B(26),C(15),T(20)      \\
	14           & Mark Taylor     & VP \& General Counsel      & D(5),EV(14),B(13),C(13),T(13)      \\
	15           & Susan Scott     & Assistant Trader           & D(16),EV(30),B(4),C(14),T(10)      \\
	16           & Sara Shackleton & VP ENA \& Senior Counsel   & D(4),EV(64),B(68),C(79),T(55)      \\
	17           & Richard Shapiro & VP Regulatory Affairs      & D(6),EV(28),B(79),C(29),T(42)      \\
	18           & Steven Kean     & VP \& Chief of Staff       & D(8),EV(19),B(60),C(21),T(29)      \\
	19           & Bill Williams   & Trader                     & D(62),EV(120),B(7),C(88),T(75)     \\
	20           & John Forney     & Manager Real Time Trading  & D(81),EV(88),B(9),C(36),T(49)      \\
	21           & Lysa Akin       & Sr Adm Asst Gov Affairs    & D(73),EV(81),B(10),C(63),T(39)     \\
	22           & Carol Clair     & Assistant General Counsel  & D(10),EV(103),B(128),C(119),T(111) \\ \hline
\end{tabular}
\label{t:top10comnet}
\end{table*}

\begin{table*}
\centering
\caption{Top 10 employees for centrality measures in Attachments Network}
\begin{tabular}{|l|l|l|l|}
	\hline
	Overall Rank & Name                      & Title                           & Centrality Measures              \\ \hline
	1            & Phillip Allen(7)\{1\}     & Managing Director Trading       & D(12),EV(1),B(2),C(1),T(1)       \\
	2            & James Steffes(13)\{7\}    & VP Government Affairs           & D(1),EV(18),B(6),C(3),T(8)       \\
	3            & Mike Grigsby(9)\{3\}      & VP Trading, ENA Gas West        & D(14),EV(2),B(4),C(2),T(3)       \\
	4            & Hunter Shively\{6\}       & VP Trading, ENA Gas Central     & D(46),EV(3),B(3),C(5),T(2)       \\
	5            & John Lavorato(1)\{10\}    & CEO, ENA                        & D(19),EV(4),B(8),C(4),T(4)       \\
	6            & Elizabeth Sager\{16\}     & VP \& Assistant General Counsel & D(8),EV(33),B(7),C(10),T(13)     \\
	7            & Susan Scott(15)           & Assistant Trader                & D(11),EV(13),B(1),C(8),T(5)      \\
	8            & Keith Holst\{19\}         & Trader                          & D(20),EV(5),B(12),C(6),T(6)      \\
	9            & Steven Kean(18)\{12\}     & VP \& Chief of Staff            & D(4),EV(14),B(25),C(9),T(11)     \\
	10           & Tana Jones(12)\{17\}      & Senior Legal Specialist         & D(2),EV(71),B(71),C(51),T(52)    \\
	11           & Richard Shapiro(17)\{14\} & VP Regulatory Affairs           & D(3),EV(22),B(36),C(12),T(19)    \\
	12           & Jeff Dasovich(6)\{13\}    & Dir State Government Affairs    & D(5),EV(28),B(24),C(17),T(23)    \\
	13           & Kevin Presto\{5\}         & VP Trading, ENA East Power      & D(44),EV(19),B(11),C(7),T(9)     \\
	14           & Matthew Lenhart\{9\}      & Analyst                         & D(18),EV(10),B(19),C(11),T(14)   \\
	15           & Barry Tycholiz\{8\}       & VP Trading, ENA Gas West        & D(16),EV(7),B(34),C(19),T(12)    \\
	16           & Williams Jason            & Trader, ENA Gas Central         & D(39),EV(8),B(15),C(24),T(7)     \\
	17           & Thomas Martin             & VP Trading, ENA Gas Texas       & D(66),EV(6),B(40),C(23),T(17)    \\
	18           & Jay Reitmeyer\{11\}       & Associate                       & D(23),EV(9),B(43),C(15),T(15)    \\
	19           & Mike Swerzbin             & VP Trading, ENA West Power      & D(117),EV(76),B(5),C(50),T(62)   \\
	20           & Mark Taylor\{2\}          & VP \& General Counsel           & D(6),EV(47),B(45),C(33),T(34)    \\
	21           & Marie Heard               & Specialist Legal                & D(7),EV(80),B(85),C(61),T(73)    \\
	22           & Sara Shackleton(16)       & VP ENA \& Senior Counsel        & D(9),EV(69),B(48),C(34),T(57)    \\
	23           & Robert Badeer             & Mgr Trading, ENA West Power     & D(67),EV(60),B(9),C(32),T(49)    \\
	24           & John Forney               & Dir Trading, ENA East Power     & D(100),EV(94),B(10),C(38),T(71)  \\
	25           & Stacy Dickson             & Senior Counsel, ENA             & D(10),EV(100),B(80),C(94),T(103) \\ \hline
\end{tabular}
\label{t:top10atnet}
\end{table*}

We calculate overall statistics and 4 measures of centrality: degree\footnote{\url{https://en.wikipedia.org/wiki/Degree\_(graph\_theory)}}, eigenvector\footnote{\url{https://en.wikipedia.org/wiki/Eigenvector\_centrality}}, betweenness\footnote{\url{https://en.wikipedia.org/wiki/Betweenness\_centrality}}, and closeness\footnote{\url{https://en.wikipedia.org/wiki/Closeness\_centrality}} for both networks. These measures are in no way exclusive but are frequently used in SNA to infer most influential people in a network. Enron's organizational charts are used throughout the analysis. The charts are a combination of previous research \cite{Agarwal2012}, EnronEmployeeInformation.csv\footnote{\url{http://ww2.amstat.org/publications/jse/jse_data_archive.html}}, documents related to Enron's legal proceedings\footnote{\url{https://www.gpo.gov/fdsys/pkg/GPO-CPRT-JCS-3-03/pdf/GPO-CPRT-JCS-3-03-3-2-8.pdf}}, Enron's emails, and information we discovered on LinkedIn\footnote{\url{https://www.linkedin.com}}.
We used Python's NetworkX module\footnote{\url{https://networkx.github.io}} for the overall network statistics and centrality measures. 
Table ~\ref{t:generalstats} shows the general statistics for both networks. 
For the Attachments Network we show statistics with the filtered and un-filtered attachments. 
We see that filtered Attachments Network as compared to Communications Network is less connected, less populated with edges, and less centralized. It has one less node ``Monika Causholli'', who doesn't have any shared attachments with other employees when filters are applied. 
Unfiltered attachments Network is significantly different from the other two. 
Most notably the clustering coefficient, characteristic path length, average number of neighbors, and network density point towards the network with more triads and more populated with edges. 
This is the result of discovering more friend of a friend relationship by connecting users via attachments shared in their email archives. 
We effectively extended the friend definition from 'someone who has direct communication with me' to 'someone who has some information shared with me'. 
In extreme case when the bulk email is sent to everyone, then everyone shares the same information, resulting in complete graph.

Table \ref{t:top10comnet} shows top 10 employees ranking in one of degree(D), eigenvector(EV), betweenness(B), closeness(C) centrality measures and unique ties(T) in Communication Network. 
We calculated the overall rank as the sum of inverse value of ranking in the centrality measure plus one for each centrality measure which is in top 10. 
The last part favors employees who have more top 10 ranking. 
It is not surprising that the most influential employees are Executives. 
Number one is John Lavorato, CEO North America. 
Liz Taylor is assistant to President \& COO Greg Whalley, who is not ranked in any top 10 measures. 
It is reasonable to assume that Liz Taylor is the proxy for Greg Whalley. 
Louise Kitchen is number three and is one of the most influential people at Enron where she pioneered the on-line trading.
Number four is Sally Beck, COO. 
And number five is Kenneth Lay, CEO of the Company. 
Overall, the list of the most influential employees underscores the importance of Trading, with 8 representatives, Legal Department with 4 representatives, and Regulatory and Government Affairs with 4 representatives. 
There are 5 non-executive level employees on the list. 
One of them, Bill Williams was implicated in the energy price fixing at Enron\footnote{\url{http://www.nytimes.com/2005/02/04/us/tapes-show-enron-arranged-plant-shutdown.html}} and was managing the largest trading group shortly before the Enron's collapse\footnote{\url{http://www.mresearch.com/pdfs/89.pdf}}. 

Table ~\ref{t:top10atnet} shows top 10 employees in one of the centrality measures in filtered Attachments Network. 
It is still dominated by high level executives but also there are more regular employees. 
There are 10 employees from Communication Network who held their position in one of the top 10 centrality measures but only three, Phillip Allen, Mike Grigsby, and John Lavorato remained in the overall top 10 ranking. 
We show Communication Network ranking in parenthesis next to the employee's name. 
Overall, traders are dominating the ranks with 14 representatives out of which 3 traders are in top 5. 
This underscores that the main Enron's business is Energy trading. 
Next there are 6 representatives from Legal Department, which shows that Enron had to address many legal issues as part of the energy trading. 
This correlates with the fact that out of 15 new top employees in Attachments Network, 11 are traders and 4 are from Legal Department. 
This again stresses the influence of Trading and Legal Departments within Enron Organization. 
This also indicates Traders and Legal Department employees had to deal with substantial document handling as part of their responsibilities. 

\begin{table*}
\centering
\caption{Top 10 gained ties in Attachments Network}
\begin{tabular}{|l|l|l|l|}
	\hline
	Name                        & Name                             & Friend of a Friend        & Ties \\ \hline
	Stacy Dickson,SnrCnsl       & Marie Heard, SnrLglSpcl          & Tana Jones, SnrLglSpcl    & 128  \\
	Stacy Dickson,SnrCnsl       & Elizabeth Sager, VP\&AsstGenCnsl & Tana Jones, SnrLglSpcl    & 108  \\
	Keith Holst,Trader          & Frank Ermis,Dir Trading          & Mike Grigsby,VP Trading   & 20   \\
	Tori Kuykendall,Mgr Trading & Jason Wolfe,Trader               & Mike Grigsby,VP Trading   & 18   \\
	Scott Hendrickson,Trader    & Judy Townsend,Trader             & Chris Germany,Mgr Trading & 16   \\
	Randal Gay,Trader           & Keith Holst,Trader               & Mike Grigsby,VP Trading   & 16   \\
	Randal Gay,Trader           & Jason Wolfe,Trader               & Mike Grigsby,VP Trading   & 15   \\
	Matt Smith,Associate        & Jason Wolfe,Trader               & Mike Grigsby,VP Trading   & 15   \\
	Matt Smith,Associate        & Randall Gay,Trader               & Mike Grigsby,VP Trading   & 15   \\
	Jay Reitmeyer, Associate    & Barry Tycholiz, VP Trading       & Mike Grigsby, VP Trading  & 14   \\ \hline
\end{tabular}
\label{t:top10gained}
\end{table*}

The rank in the curly brackets next to the employee name in the Table \ref{t:top10atnet} shows the overall ranking of the employee in the unfiltered Attachments Network. 
We see that employees move in both direction of the overall ranking in 'filtered' as compared to 'unfiltered' networks and 2 employees Louise Kitchen\{4\} and Kevin Presto\{5\} have been removed from any top 10 ranking. 

Besides affecting top 10 employees, the different definition of ties had, as expected, substantial impact on the overall number of ties in the network. 
There are 388 new ties and 704 lost ties in the Attachments as compared to the Communication Network. 
We reviewed top 10 new ties in Table ~\ref{t:top10gained}. The 'Friend of a Friend' column shows a 'friend' who contributed the most shared attachments that created a tie between two employees.
Employees have their ties due to a friend of a friend relationship where the friend has the same functional position as employees. 
For instance, Stacy Dickson and Marie Heard are both from Legal Department, ENA. 
Their tie, or shared attachment was created by Tana Jones, who sent a document to both of them, i.e. they both were on the To, Cc, or Bcc list. 

Analysis of the lost ties shows that in those cases employees had fewer shared attachments that were filtered out or no shared attachments at all. 
Also, out of 704 lost ties, 593 had the communication frequency less or equal to 5 emails and 353 out 388 new ties had less or equal to 5 shared attachments. 
Consequently, if the lower bound threshold on the frequency of ties is set to 5 then there will be fewer lost and gained ties.

The analysis above is a guesstimate in nature. I.e. based on centrality measures of the extracted shared attachments network, we rated top 10 most influential employees and attempted to corroborate our findings with the Enron's organizational charts and some on-line news stories. The extracted network is based on the information sharing and reflects inter and intra interactions within functional groups and does not necessarily overlap with the organization charts. This type of analysis can be used by sociologists, anthropologists, or managers to improve efficiency of the communication within as well as outside of a company. The gained ties analysis can be used to discover hidden relationships, which can not be discovered with the communication network. This is the use case we covered in section \ref{s:netextract} when we discussed edge direction in the shared attachments network. Other type of analysis could be similar to~\cite{Wang2012} where a Jaccard similarity matrix is generated and then clusters of user groups are derived from this matrix. According to homophily principal~\cite{Mcpherson2001} a contact between similar people occurs at a higher rate than amongst dissimilar people. Since the focus of the shared attachments network is on the information sharing, the similarity and clustering approach may produce a valuable insight into functional, organizational, and social group interactions. 

\subsection{K-Nearest Neighbor}
\label{s:neighbor}

We are analyzing similarity of some Enron employees based on the k-nearest neighbor algorithm. The idea is that each attachment's SHA1 can be viewed as a word. Then if we have a 'bag of words' for each employee, we could derive a list of employees similar to an employee of interest by building the nearest neighbor model on all employees dataset and querying the model for the employee. We use Graphlab\footnote{\url{https://turi.com/learn/userguide/index.html}} Python's module to query for similarity list of five employees. Listing \ref{l:neighbor} shows the Python code example to generate the nearest neighbor list. 

\begin{lstlisting}[language=Python, caption=Python k-nearest neighbor example, label=l:neighbor]
import graphlab as gl
data = gl.SFrame.read_csv('./shared.txt', 
  delimiter=',', header = True)
data['words'] = gl.text_analytics.count_words(
  data['attachments'])
model = gl.nearest_neighbors.create(data, 
  features=['words'], label='name')
model.query(data[data['name'] == 'Kenneth Lay'],k=6)
\end{lstlisting}

We generated similarity lists for five Enron's employees: Kenneth Lay, Enron Chairman \& CEO; Jeff Skilling, Enron President \& COO, John Lavorato, Enron Americas COO, is ranked number one in the Communication Network; Phillip Allen, Managing Director Trading, is ranked number one in the Attachments Network; Stacy Dickson, ENA Attorney, is the top gained tie. The lists are shown in Table ~\ref{t:similar}.

\begin{table*}
	\centering
	\caption{Similarity lists based on the nearest neighbor model. The name in parenthesis is employee's manager. }
	\begin{tabular}{|l|l|l|}
		\hline
		Name             & Title                                               & Reporting to                                         \\ \hline
		Kenneth Lay      & Enron Chairman \& CEO                               & Board of Directors                                   \\
		Rosalee Fleming  & Enron Chairman's secretary                          & Kenneth Lay                                          \\
		Greg Whalley     & EWS President \& COO                                & Mike Frevert EWS Chairman \& CEO(Jeff Skilling)      \\
		James Derrick    & Executive VP \& General Counsel                     & Jeff Skilling                                        \\
		Steven Kean      & Executive VP \& Chief of Staff                      & Jeff Skilling                                        \\
		Jeff Skilling    & Enron President \& COO                              & Kenneth Lay                                          \\ \hline\hline
		Jeff Skilling    & Enron President \& COO                              & Kenneth Lay                                          \\
		Rick Buy         & Executive VP \& Chief Risk Officer                  & Jeff Skilling                                        \\
		James Derick     & Executive VP \& General Counsel                     & Jeff Skilling                                        \\
		Kenneth Lay      & Enron Chairman \& COO                               & Board of Directors                                   \\
		Greg Whalley     & EWS President \& COO                                & Mark Frevert, EWS Chairman \& CEO(Jeff Skilling)     \\
		David Delainey   & EA President \& COO                                 & Greg Whalley                                         \\ \hline\hline
		John Lavorato    & EA COO                                              & David Delainey                                       \\
		Louise Kitchen   & EN President \& CEO                                 & Greg Whalley                                         \\
		David Delainey   & EA President \& COO                                 & Greg Whalley                                         \\
		Greg Whalley     & EWS President \& COO                                & Mark Frevert, EWS Chairman \& CEO(Jeff Skilling)     \\
		Kevin Presto     & ENA EP VP                                           & John Lavorato                                        \\
		Jeffrey Shankman & EGM COO                                             & Mike McConnell EGM President \& CEO(Greg Whalley)    \\ \hline\hline
		Phillip Allen    & ENA GW Managing Director Trading                    & John Lavorato                                        \\
		Mike Grigsby     & ENA GW VP Trading                                   & Phillip Allen                                        \\
		Keith Holst      & ENA GW Director Trading                             & Vince Kaminski EWS Managing Director (John Lavorato) \\
		Matt Smith       & ENA GW Associate                                    & Not Available                                        \\
		Matthew Lenhart  & ENA GW Analyst                                      & Mike Grigsby                                         \\
		Jane Tholt       & ENA GW Director Trading                             & Mike Grigsby                                         \\ \hline\hline
		Stacy Dickson    & ENA Attorney                                        & Jeff Hodge ENA VP \& AGC(Mark Haedicke)              \\
		Tana Jones       & Net Works Financial Trading Senior Legal Specialist & Mark Taylor VP \& GC(Mark Haedicke)                  \\
		Marie Heard      & ENA Legal Specialist                                & Not Available                                        \\
		Elizabeth Sager  & ENA Power Trading AGC                               & Mark Haedicke EWS Managing Director \& GC            \\
		Mark Taylor      & Net Works Financial Trading VP \& GC                & Mark Haedicke EWS Managing Director \& GC            \\
		Debra Perlingier & ENA Senior Legal Specialist                         & Jeff Hodge ENA VP \& AGC(Mark Haedicke)              \\ \hline
	\end{tabular}
\label{t:similar}
\end{table*}

We see that within each similarity group employees are either direct report of the top employee in the group, the manager of the employee, the peer, the descendant in the same tree branch of the organizational chart, or appear to be part of a functional group. There are some interesting points about the lists. Rosalee Fleming is not listed in any of the organizational charts that we used for the title reference. We found a reference to her title in one of Enron's email. Her degree centrality measure ranks number 17 in Kenneth Lay's ego communication network. Clearly, as Kenneth Lay's secretary she should had handled information exchange between Kenneth Lay and other employees. And the data fed from the attachments sharing network into the nearest neighbor model shows just that. Other points are on Matt Smith and Marie Heard. Neither of them has the information available regarding their manager. The nearest neighbor model accurately predicts their similarity to ENA GW and Legal department respectively judging by other employees title in the similarity groups. We find these results encouraging and validating our approach to extracting social network via email's shared attachments.

\subsection{K-Means Clustering}
\label{s:clustering}

As another way of evaluating the attachments network we classify employees with k-means clustering algorithm.
K-means is one of the simplest and mostly used unsupervised learning algorithms that solves the clustering problem. We run the algorithm on weighted Jaccard distance derived from shared attachments between pairs of employees. K-means takes as an input parameter the number of clusters. We assume that the information exchange is higher within an organizational unit whether it is the functional team or the department. Since the functional team information is not available, we guesstimate the number of clusters as the average number of employees in a department. Based on Enron organizational charts, the average department size in Enron is 10 employees. Since there are 150 core employees in the Enron dataset, we set the number of clusters to 15. We use SciKit\footnote{\url{http://scikit-learn.org/stable/modules/generated/sklearn.cluster.KMeans.html}} Python's clustering module. Listing \ref{l:clusters} shows the code example to generate the clusters. Table \ref{t:clusters} shows the result of the clustering. Within each cluster we group employees by their respective department.

\begin{table*}
	\centering
	\caption{K-means clustering results for shared attachments network. Employees are grouped by the department within the cluster.}
	\begin{tabular}{|l|l|l|l|}
		\hline
		Cl. \# & Cl. size & Department                        & \# employees \\ \hline
		1      & 9        & ENA West Power Real Time          & 9            \\ \hline
		2      & 8        & ENA Gas Central                   & 2            \\
		       &          & ENA Gas East                      & 4            \\
		       &          & ENA Gas Texas                     & 1            \\
		       &          & Energy Operations                 & 1            \\ \hline
		3      & 40       & EES                               & 1            \\
		       &          & ENA East Power                    & 4            \\
		       &          & ENA Gas Central                   & 7            \\
		       &          & ENA Gas East                      & 4            \\
		       &          & ENA Gas Financial                 & 2            \\
		       &          & ENA Gas Texas                     & 1            \\
		       &          & ENA Gas West                      & 2            \\
		       &          & ENA Legal                         & 5            \\
		       &          & ENA West Power                    & 5            \\
		       &          & ETS                               & 1            \\
		       &          & EWS                               & 3            \\
		       &          & Energy Operations                 & 4            \\
		       &          & Regulatory and Government Affairs & 1            \\ \hline
		4      & 11       & ENA East Power                    & 1            \\
		       &          & ENA Gas Central                   & 1            \\
		       &          & ENA Gas East                      & 4            \\
		       &          & ENA Gas Financial                 & 3            \\
		       &          & ENA Gas Texas                     & 2            \\ \hline
		5      & 6        & ENA Gas West                      & 4            \\
		       &          & ENA Legal                         & 2            \\ \hline
		6      & 5        & ETS                               & 5            \\ \hline
		7      & 6        & ENA East Power                    & 4            \\
		       &          & ENA Legal                         & 1            \\
		       &          & Energy Operations                 & 1            \\ \hline
		8      & 11       & ENA Gas West                      & 1            \\
		       &          & ETS                               & 10           \\ \hline
		9      & 13       & ENA Legal                         & 1            \\
		       &          & ETS                               & 1            \\
		       &          & EWS                               & 6            \\
		       &          & Enron                             & 5            \\ \hline
		10     & 6        & ENA West Power                    & 5            \\
		       &          & ENA West Power Real Time          & 1            \\ \hline
		11     & 5        & ENA East Power                    & 5            \\ \hline
		12     & 6        & ENA Legal                         & 6            \\ \hline
		13     & 13       & ENA Gas Texas                     & 1            \\
		       &          & ENA Gas West                      & 12           \\ \hline
		14     & 4        & ENA East Power                    & 4            \\ \hline
		15     & 6        & ENA Legal                         & 1            \\
		       &          & Enron                             & 1            \\
		       &          & Regulatory and Government Affairs & 4            \\ \hline
	\end{tabular}
	\label{t:clusters}
\end{table*}

We see that employees in clusters 1, 6, 11, 12, and 14 are entirely within the departmental boundaries. Cluster 9 consists of top level corporate executives, including Kenneth Lay(CEO) and Jeffrey Skilling(President). Clusters 8, 10, and 13 have only one employee who is not in the same department with the rest of the employees. Cluster 15 also has only one employee who is not in the same department as other employees. The employee from Enron department is Steven Kean, Exec VP\&Chief of Staff. His responsibilities included Regulatory\&Government Affairs.  Clusters 2 and 4 have only one employee who is not in one of the ENA Gas departments. Clusters 5 and 7 have two employees who are not in the same department as the rest of the employee. Cluster 3 has the highest number of employees from 13 departments. Majority of employees in this cluster are from ENA Gas departments. 4 out of 5 Legal department representative are from Financial and Gas trading groups from within the Legal organization. While we see that the clustering has grouped many employees by their respective departments, it would not be correct to generalize this result as ability of this approach to predict organizational units. The shared attachments network reflects information interactions between employees and the clustering shows information exchange between functional units that happened to overlap in some cases with organizational units or departments. We can conclude that many departments by end large have majority of communication either within the department or the subset of the department boundary. Top level corporate management has its own clique, perhaps highlighting lack of top-down information interaction within the organization. The largest cluster number 3 is the most diverse in terms of inter-department information exchange and judging by the number of representative from ENA Gas is mostly involved in Gas trading with other departments providing necessary support.

\begin{lstlisting}[language=Python, caption=Python k-means clustering example, label=l:clusters]
import graphlab as gl
from sklearn.cluster import KMeans
# attachments is a dictionary keyed by 
# the employee name with the value 
# containing the list of attachments's SHA1
distance = list()
sorted_names = sorted(attachments.keys())

for name1 in sorted_names:
  row = list()
  for name2 in sorted_names:
    d = gl.toolkits.distances.weighted_jaccard(
      dict(attachments[name1]), 
      dict(attachments[name2]))
    row.extend([d])
  distance.extend([row])

model = KMeans(n_clusters=15, random_state = 0)
model.fit(distance)
clusters = defaultdict(list)
for r in range(0, len(model.labels_)):
clusters[mode.labels_[r]].extend(
  [sorted_names[r]])
# clusters is a dictionary keyed by 
# the cluster number with the value 
# containing the list of employee's name
\end{lstlisting}

\section{Conclusion}
\label{s:conclusion}

We linked EDRMV2 Enron email dataset containing attachments with Shetty\&Adibi Enron email dataset, which has been extensively used in the previous research. 
We then extracted two social networks. 
First, with nodes representing employees and ties representing communication between employees. 
Second, with nodes representing employees and ties representing email attachments shared between employees. 
We suggested a set of rules to filter out TRAM attachments like common logos, signatures, or Internet trend images, which could be commonly shared between people and do not represent a valuable tie. 
Four centrality measures - degree, eigenvector, betweenness, and closeness were calculated and the resulting ranked list of employees in top 10 of each measure was analyzed for each network. 
Extracting the social network from shared attachments could be complimentary to the generally used communication network and may discover more key influential people in the network and help to infer friend of a friend relationship. We demonstrated that the nearest neighbor model accurately predicts similarity between employees, which is corroborated by employee's position in the organizational charts. K-means clustering shows groups with users in each group related to each other via their functional responsibilities. 
We see two main challenges in extracting the shared attachments network. 
First is defining the quality ties, and second is defining the strength of the tie. 
Both of these topics are left for the future research.
In this paper we demonstrated viability of extracting social network from email shared attachments.   
Because of the privacy concerns it is notoriously difficult to obtain an email corpus for an analysis. Consequently, our evaluation and ability to generalize is constrained by the available data. We are hoping that this paper will motivate large email providers to apply our approach in their SNA research.

\bibliographystyle{ACM-Reference-Format}
\bibliography{paper}


\begin{thebibliography}{00}


\ifx \showCODEN    \undefined \def \showCODEN     #1{\unskip}     \fi
\ifx \showDOI      \undefined \def \showDOI       #1{{\tt DOI:}\penalty0{#1}\ }
  \fi
\ifx \showISBNx    \undefined \def \showISBNx     #1{\unskip}     \fi
\ifx \showISBNxiii \undefined \def \showISBNxiii  #1{\unskip}     \fi
\ifx \showISSN     \undefined \def \showISSN      #1{\unskip}     \fi
\ifx \showLCCN     \undefined \def \showLCCN      #1{\unskip}     \fi
\ifx \shownote     \undefined \def \shownote      #1{#1}          \fi
\ifx \showarticletitle \undefined \def \showarticletitle #1{#1}   \fi
\ifx \showURL      \undefined \def \showURL       #1{#1}          \fi
\providecommand\bibfield[2]{#2}
\providecommand\bibinfo[2]{#2}
\providecommand\natexlab[1]{#1}
\providecommand\showeprint[2][]{arXiv:#2}

\bibitem[\protect\citeauthoryear{Agarwal, Omuya, Harnly, and Rambow}{Agarwal
  et~al\mbox{.}}{2012}]%
        {Agarwal2012}
\bibfield{author}{\bibinfo{person}{Apoorv Agarwal}, \bibinfo{person}{Adinoyi
  Omuya}, \bibinfo{person}{Aaron Harnly}, {and} \bibinfo{person}{Owen Rambow}.}
  \bibinfo{year}{2012}\natexlab{}.
\newblock \showarticletitle{{A Comprehensive Gold Standard for the Enron
  Organizational Hierarchy}}. In \bibinfo{booktitle}{{\em Proceedings of the
  50th Annual Meeting of the Association for Computational Linguistics}}.
  \bibinfo{address}{Jeju, Republic of Korea}, \bibinfo{pages}{161--165}.
\newblock
\showISBNx{9781937284251}


\bibitem[\protect\citeauthoryear{Bartlett, Kotrlik, and Higgins}{Bartlett
  et~al\mbox{.}}{2001}]%
        {Bertlett2001}
\bibfield{author}{\bibinfo{person}{James~E. Bartlett}, \bibinfo{person}{Joe~W.
  Kotrlik}, {and} \bibinfo{person}{Chadwick~C Higgins}.}
  \bibinfo{year}{2001}\natexlab{}.
\newblock \showarticletitle{{Organizational Research: Determining Appropriate
  Sample Size in Survey Research Appropriate Sample Size in Survey Research}}.
\newblock \bibinfo{journal}{{\em Information Technology, Learning and
  Performance Journal\/}} \bibinfo{volume}{19}, \bibinfo{number}{1}
  (\bibinfo{year}{2001}), \bibinfo{pages}{43--50}.
\newblock
\showISBNx{15351556}
\showISSN{15351556}
\showDOI{%
\url{https://doi.org/10.1109/LPT.2009.2020494}}


\bibitem[\protect\citeauthoryear{Bird, Gourley, Devanbu, and Gertz}{Bird
  et~al\mbox{.}}{2006}]%
        {Bird2006}
\bibfield{author}{\bibinfo{person}{Christian Bird}, \bibinfo{person}{Alex
  Gourley}, \bibinfo{person}{Prem Devanbu}, {and} \bibinfo{person}{Michael
  Gertz}.} \bibinfo{year}{2006}\natexlab{}.
\newblock \showarticletitle{{Mining Email Social Networks}}. In
  \bibinfo{booktitle}{{\em MSR'06}}. \bibinfo{address}{Shanghai, China},
  \bibinfo{pages}{137--143}.
\newblock
\showISBNx{159593085X}
\showDOI{%
\url{https://doi.org/10.1145/1137983.1138016}}


\bibitem[\protect\citeauthoryear{Choudhury, Mason, Hofman, and Watts}{Choudhury
  et~al\mbox{.}}{2010}]%
        {Choudhury2010}
\bibfield{author}{\bibinfo{person}{Munmun~De Choudhury},
  \bibinfo{person}{Winter~A Mason}, \bibinfo{person}{Jake~M Hofman}, {and}
  \bibinfo{person}{Duncan~J Watts}.} \bibinfo{year}{2010}\natexlab{}.
\newblock \showarticletitle{{Inferring Relevant Social Networks from
  Interpersonal Communication}}. In \bibinfo{booktitle}{{\em WWW 2010}}.
  \bibinfo{address}{Raleigh, North Carolina, USA}, \bibinfo{pages}{301--310}.
\newblock
\showISBNx{9781605587998}


\bibitem[\protect\citeauthoryear{Diesner, Frantz, and Carley}{Diesner
  et~al\mbox{.}}{2006}]%
        {Diesner2006}
\bibfield{author}{\bibinfo{person}{Jana Diesner}, \bibinfo{person}{L.~Terrill
  Frantz}, {and} \bibinfo{person}{M.~Kathleen Carley}.}
  \bibinfo{year}{2006}\natexlab{}.
\newblock \showarticletitle{{Communication Networks from the Enron Email Corpus
  `` It ' s Always About the People . Enron is no Different ''}}.
\newblock \bibinfo{journal}{{\em Computational {\&} Mathematical Organization
  Theory\/}} \bibinfo{volume}{11}, \bibinfo{number}{3} (\bibinfo{year}{2006}),
  \bibinfo{pages}{201--228}.
\newblock
\showDOI{%
\url{https://doi.org/10.1007/s10588-005-5377-0}}


\bibitem[\protect\citeauthoryear{Dunbar}{Dunbar}{1998}]%
        {Dunb}
\bibfield{author}{\bibinfo{person}{Robin~I.M. Dunbar}.}
  \bibinfo{year}{1998}\natexlab{}.
\newblock \showarticletitle{{The Social Brain Hypothesis}}.
\newblock \bibinfo{journal}{{\em Evolutionary Antropology\/}}
  (\bibinfo{year}{1998}), \bibinfo{pages}{178--190}.
\newblock


\bibitem[\protect\citeauthoryear{Dunbar}{Dunbar}{2011}]%
        {Dunbar2011}
\bibfield{author}{\bibinfo{person}{Robin~I.M. Dunbar}.}
  \bibinfo{year}{2011}\natexlab{}.
\newblock \bibinfo{title}{{How Many {"Friends"} Can You Really Have?}}
\newblock   (\bibinfo{year}{2011}).
\newblock
\showURL{%
\url{http://spectrum.ieee.org/telecom/internet/how-many-friends-can-you-really-have}}


\bibitem[\protect\citeauthoryear{Dunbar}{Dunbar}{1993}]%
        {Dunbar1993}
\bibfield{author}{\bibinfo{person}{R~I~M Dunbar}.}
  \bibinfo{year}{1993}\natexlab{}.
\newblock \showarticletitle{{Co-Evolution of Neocortex Size, Group Size and
  Language in Humans}}.
\newblock   \bibinfo{volume}{16} (\bibinfo{year}{1993}),
  \bibinfo{pages}{1--27}.
\newblock
\showURL{%
\url{http://www.bbsonline.org/documents/a/00/00/05/65/bbs000005...}}


\bibitem[\protect\citeauthoryear{Easlake, Huawei, and Nansen}{Easlake
  et~al\mbox{.}}{2011}]%
        {Easlake2011}
\bibfield{author}{\bibinfo{person}{D.~3rd Easlake}, \bibinfo{person}{Huawei},
  {and} \bibinfo{person}{T. Nansen}.} \bibinfo{year}{2011}\natexlab{}.
\newblock \showarticletitle{{rfc 6234: US Secure Hash Algorithms}}.
\newblock \bibinfo{journal}{{\em Internet Engineering Task Force\/}}
  (\bibinfo{year}{2011}).
\newblock
\showURL{%
\url{https://tools.ietf.org/html/rfc6234}}


\bibitem[\protect\citeauthoryear{Freed and Borenstein}{Freed and
  Borenstein}{1996}]%
        {Freed1996a}
\bibfield{author}{\bibinfo{person}{N Freed} {and} \bibinfo{person}{N
  Borenstein}.} \bibinfo{year}{1996}\natexlab{}.
\newblock \showarticletitle{{RFC 2045: Multipurpose Internet Mail Extensions
  (MIME) Part One: Format of Internet Message Bodies.}}
\newblock \bibinfo{journal}{{\em Network Working Group\/}}
  \bibinfo{number}{November} (\bibinfo{year}{1996}).
\newblock


\bibitem[\protect\citeauthoryear{Gilbert and Karahalios}{Gilbert and
  Karahalios}{2009}]%
        {Gilbert2009}
\bibfield{author}{\bibinfo{person}{Eric Gilbert} {and} \bibinfo{person}{Karrie
  Karahalios}.} \bibinfo{year}{2009}\natexlab{}.
\newblock \showarticletitle{{Predicting tie strength with social media}}. In
  \bibinfo{booktitle}{{\em Proceedings of the 27th international conference on
  Human factors in computing systems}}. \bibinfo{pages}{211--220}.
\newblock
\showISBNx{9781605582467}
\showISSN{01672738}
\showDOI{%
\url{https://doi.org/10.1145/1518701.1518736}}


\bibitem[\protect\citeauthoryear{Granovetter}{Granovetter}{1973}]%
        {Granovetter1973}
\bibfield{author}{\bibinfo{person}{Mark~S. Granovetter}.}
  \bibinfo{year}{1973}\natexlab{}.
\newblock \showarticletitle{{The Strength of Weak Ties}}.
\newblock \bibinfo{journal}{{\it Amer. J. Sociology}} \bibinfo{volume}{78},
  \bibinfo{number}{6 (May 1973)} (\bibinfo{year}{1973}),
  \bibinfo{pages}{1360--1380}.
\newblock


\bibitem[\protect\citeauthoryear{Guillaume and Latapy}{Guillaume and
  Latapy}{2006}]%
        {Guillaume2006}
\bibfield{author}{\bibinfo{person}{Jean-Loup Guillaume} {and}
  \bibinfo{person}{Matthieu Latapy}.} \bibinfo{year}{2006}\natexlab{}.
\newblock \showarticletitle{{Bipartite graphs as models of complex networks}}.
\newblock \bibinfo{journal}{{\em Physica A: Statistical Mechanics and its
  Applications\/}} \bibinfo{volume}{371}, \bibinfo{number}{2}
  (\bibinfo{year}{2006}), \bibinfo{pages}{795--813}.
\newblock
\showISBNx{0378-4371}
\showISSN{03784371}
\showDOI{%
\url{https://doi.org/10.1016/j.physa.2006.04.047}}


\bibitem[\protect\citeauthoryear{Gupte and Eliassi-Rad}{Gupte and
  Eliassi-Rad}{2012}]%
        {Gupte2012}
\bibfield{author}{\bibinfo{person}{Mangesh Gupte} {and} \bibinfo{person}{Tina
  Eliassi-Rad}.} \bibinfo{year}{2012}\natexlab{}.
\newblock \showarticletitle{{Measuring tie strength in implicit social
  networks}}. In \bibinfo{booktitle}{{\em Proceedings of the 3rd Annual ACM Web
  Science Conference on - WebSci '12}}. \bibinfo{pages}{109--118}.
\newblock
\showISBNx{9781450312288}
\showISSN{9781450312288}
\showDOI{%
\url{https://doi.org/10.1145/2380718.2380734}}
\showeprint[arxiv]{1112.2774}


\bibitem[\protect\citeauthoryear{Hill and Dunbar}{Hill and Dunbar}{2003}]%
        {Hill2003}
\bibfield{author}{\bibinfo{person}{R.~A. Hill} {and}
  \bibinfo{person}{Robin~I.M. Dunbar}.} \bibinfo{year}{2003}\natexlab{}.
\newblock \showarticletitle{{Social network size in humans}}.
\newblock \bibinfo{journal}{{\em Human Nature\/}} \bibinfo{volume}{14},
  \bibinfo{number}{1} (\bibinfo{year}{2003}), \bibinfo{pages}{53--72}.
\newblock
\showISSN{1045-6767}
\showDOI{%
\url{https://doi.org/10.1007/s12110-003-1016-y}}


\bibitem[\protect\citeauthoryear{Kaye, Khatami, Metz, and Proulx}{Kaye
  et~al\mbox{.}}{2014}]%
        {Kaye2014}
\bibfield{author}{\bibinfo{person}{T Kaye}, \bibinfo{person}{D Khatami},
  \bibinfo{person}{D Metz}, {and} \bibinfo{person}{E Proulx}.}
  \bibinfo{year}{2014}\natexlab{}.
\newblock \bibinfo{title}{{QUANTIFYING AND COMPARING CENTRALITY MEASURES FOR
  NETWORK INDIVIDUALS AS APPLIED TO THE ENRON CORPUS}}.
\newblock   (\bibinfo{year}{2014}).
\newblock
\showURL{%
\url{https://www.siam.org/students/siuro/vol7/S01320.pdf}}


\bibitem[\protect\citeauthoryear{Laclav{\'{i}}k, Dlugolinsk{\'{y}}, Kvassay,
  and Hluch{\'{y}}}{Laclav{\'{i}}k et~al\mbox{.}}{2011}]%
        {Laclavik2011}
\bibfield{author}{\bibinfo{person}{Michal Laclav{\'{i}}k},
  \bibinfo{person}{{\v{S}}tefan Dlugolinsk{\'{y}}}, \bibinfo{person}{Marcel
  Kvassay}, {and} \bibinfo{person}{Ladislav Hluch{\'{y}}}.}
  \bibinfo{year}{2011}\natexlab{}.
\newblock \showarticletitle{{Email social network extraction and search}}. In
  \bibinfo{booktitle}{{\em Proceedings - 2011 IEEE/WIC/ACM International Joint
  Conferences on Web Intelligence and Intelligent Agent Technology - Workshops,
  WI-IAT 2011}}, Vol.~\bibinfo{volume}{3}. \bibinfo{pages}{373--376}.
\newblock
\showISBNx{9780769545134}
\showDOI{%
\url{https://doi.org/10.1109/WI-IAT.2011.30}}


\bibitem[\protect\citeauthoryear{Laclav{\'{i}}k and
  {\v{S}}eleng}{Laclav{\'{i}}k and {\v{S}}eleng}{2012}]%
        {Laclavik2012}
\bibfield{author}{\bibinfo{person}{Michal Laclav{\'{i}}k} {and}
  \bibinfo{person}{Martin {\v{S}}eleng}.} \bibinfo{year}{2012}\natexlab{}.
\newblock \showarticletitle{{Emails as Graph : Relation Discovery in Email
  Archive}}. In \bibinfo{booktitle}{{\em Www 2012 - Companion}}.
  \bibinfo{address}{Lyon, France}, \bibinfo{pages}{841--846}.
\newblock
\showISBNx{9781450312301}
\showDOI{%
\url{https://doi.org/10.1145/2187980.2188210}}


\bibitem[\protect\citeauthoryear{Lam and Yeung}{Lam and Yeung}{2007}]%
        {Lam2007}
\bibfield{author}{\bibinfo{person}{H.-Y. Lam} {and} \bibinfo{person}{D.-Y.
  Yeung}.} \bibinfo{year}{2007}\natexlab{}.
\newblock \showarticletitle{{A Learning approach to spam detection based on
  social networks}}. In \bibinfo{booktitle}{{\em 4th conference on Email and
  anti-spam (CEAS 2007)}}. \bibinfo{pages}{81--89}.
\newblock


\bibitem[\protect\citeauthoryear{McCallum, Wang, and Corrada-Emmanuel}{McCallum
  et~al\mbox{.}}{2007}]%
        {McCallum2007}
\bibfield{author}{\bibinfo{person}{Andrew McCallum}, \bibinfo{person}{Xuerui
  Wang}, {and} \bibinfo{person}{Andr{\'{e}}s Corrada-Emmanuel}.}
  \bibinfo{year}{2007}\natexlab{}.
\newblock \showarticletitle{{Topic and role discovery in social networks with
  experiments on enron and academic email}}.
\newblock \bibinfo{journal}{{\em Journal of Artificial Intelligence
  Research\/}}  \bibinfo{volume}{30} (\bibinfo{year}{2007}),
  \bibinfo{pages}{249--272}.
\newblock
\showISBNx{1076-9757}
\showISSN{10769757}
\showDOI{%
\url{https://doi.org/10.1613/jair.2229}}
\showeprint[arxiv]{arXiv:1203.5753v5}


\bibitem[\protect\citeauthoryear{Mcpherson, Smith-lovin, and Cook}{Mcpherson
  et~al\mbox{.}}{2001}]%
        {Mcpherson2001}
\bibfield{author}{\bibinfo{person}{Miller Mcpherson}, \bibinfo{person}{Lynn
  Smith-lovin}, {and} \bibinfo{person}{James~M Cook}.}
  \bibinfo{year}{2001}\natexlab{}.
\newblock \showarticletitle{{BIRDS OF A FEATHER : Homophily in Social
  Networks}}.
\newblock \bibinfo{journal}{{\em Annual Review of Sociology\/}}
  \bibinfo{volume}{27} (\bibinfo{year}{2001}), \bibinfo{pages}{415--444}.
\newblock
\showISBNx{03600572}
\showISSN{0360-0572}
\showDOI{%
\url{https://doi.org/10.1146/annurev.soc.27.1.415}}


\bibitem[\protect\citeauthoryear{Rowe}{Rowe}{2006}]%
        {Duczynski2015}
\bibfield{author}{\bibinfo{person}{Ryan Rowe}.}
  \bibinfo{year}{2006}\natexlab{}.
\newblock \showarticletitle{{Social Hierarchy Detection through Email Network
  Analysis}}.
\newblock \bibinfo{journal}{{\em $\backslash$\/}} (\bibinfo{year}{2006}),
  \bibinfo{pages}{109--117}.
\newblock
\showISBNx{978-1-59593-848-0}
\showDOI{%
\url{https://doi.org/10.1145/1348549.1348562}}


\bibitem[\protect\citeauthoryear{Shetty and Adibi}{Shetty and Adibi}{2004}]%
        {Shetty2004}
\bibfield{author}{\bibinfo{person}{Jitesh Shetty} {and} \bibinfo{person}{Jafar
  Adibi}.} \bibinfo{year}{2004}\natexlab{}.
\newblock \showarticletitle{{The Enron Email Dataset - Database Schema and
  Brief Statistical Report}}.
\newblock \bibinfo{journal}{{\em Information Sciences Institute Technical
  Report,\/}} (\bibinfo{year}{2004}).
\newblock
\showURL{%
\url{http://citeseerx.ist.psu.edu/viewdoc/download?doi=10.1.1.296.9477}}


\bibitem[\protect\citeauthoryear{Shetty and Adibi}{Shetty and Adibi}{2005}]%
        {Shetty2005}
\bibfield{author}{\bibinfo{person}{Jitesh Shetty} {and} \bibinfo{person}{J
  Adibi}.} \bibinfo{year}{2005}\natexlab{}.
\newblock \showarticletitle{{Discovering Important Nodes through Graph Entropy:
  The Case of Enron Email Database}}. In \bibinfo{booktitle}{{\em Proceedings
  of the 3rd International Workshop on Link Discovery}}.
  \bibinfo{address}{Chicago, Illinois}, \bibinfo{pages}{74--81}.
\newblock
\showISBNx{1595932151}
\showDOI{%
\url{https://doi.org/10.1145/1134271.1134282}}


\bibitem[\protect\citeauthoryear{Tang, Barbier, Liu, and Zhang}{Tang
  et~al\mbox{.}}{2010}]%
        {Tang2010}
\bibfield{author}{\bibinfo{person}{Lei Tang}, \bibinfo{person}{Geoffrey
  Barbier}, \bibinfo{person}{Huan Liu}, {and} \bibinfo{person}{Jianping
  Zhang}.} \bibinfo{year}{2010}\natexlab{}.
\newblock \showarticletitle{{A social network analysis approach to detecting
  suspicious online financial activities}}. In \bibinfo{booktitle}{{\em Lecture
  Notes in Computer Science (including subseries Lecture Notes in Artificial
  Intelligence and Lecture Notes in Bioinformatics)}},
  Vol.~\bibinfo{volume}{6007 LNCS}. \bibinfo{publisher}{Springer-Verlag Berlin
  Heidelberg}, \bibinfo{pages}{390--397}.
\newblock
\showISBNx{3642120784}
\showISSN{03029743}
\showDOI{%
\url{https://doi.org/10.1007/978-3-642-12079-4_49}}


\bibitem[\protect\citeauthoryear{Wang, Xu, and Wang}{Wang
  et~al\mbox{.}}{2012}]%
        {Wang2012}
\bibfield{author}{\bibinfo{person}{Feng Wang}, \bibinfo{person}{Kuai Xu}, {and}
  \bibinfo{person}{Haiyan Wang}.} \bibinfo{year}{2012}\natexlab{}.
\newblock \showarticletitle{{Discovering shared interests in online social
  networks}}. In \bibinfo{booktitle}{{\em Proceedings - 32nd IEEE International
  Conference on Distributed Computing Systems Workshops, ICDCSW 2012}}.
  \bibinfo{pages}{163--168}.
\newblock
\showISBNx{978-1-4673-1423-7}
\showDOI{%
\url{https://doi.org/10.1109/ICDCSW.2012.15}}


\bibitem[\protect\citeauthoryear{Wang, Jheng, Tsai, and Tang}{Wang
  et~al\mbox{.}}{2011}]%
        {Wang2011}
\bibfield{author}{\bibinfo{person}{Min~Feng Wang}, \bibinfo{person}{Sie~Long
  Jheng}, \bibinfo{person}{Meng~Feng Tsai}, {and} \bibinfo{person}{Cheng~Hsien
  Tang}.} \bibinfo{year}{2011}\natexlab{}.
\newblock \showarticletitle{{Enterprise email classification based on social
  network features}}. In \bibinfo{booktitle}{{\em Proceedings - 2011
  International Conference on Advances in Social Networks Analysis and Mining,
  ASONAM 2011}}. \bibinfo{pages}{532--536}.
\newblock
\showISBNx{9780769543758}
\showDOI{%
\url{https://doi.org/10.1109/ASONAM.2011.89}}


\bibitem[\protect\citeauthoryear{Yelupula and Ramaswamy}{Yelupula and
  Ramaswamy}{2008}]%
        {Yelupula2008}
\bibfield{author}{\bibinfo{person}{K Yelupula} {and} \bibinfo{person}{Srini
  Ramaswamy}.} \bibinfo{year}{2008}\natexlab{}.
\newblock \showarticletitle{{Social network analysis for email
  classification}}. In \bibinfo{booktitle}{{\em Proceedings of the 46th Annual
  Southeast Regional Conference on XX - ACM-SE 46}}. \bibinfo{pages}{469}.
\newblock
\showISBNx{9781605581057}
\showDOI{%
\url{https://doi.org/10.1145/1593105.1593229}}


\bibitem[\protect\citeauthoryear{Zhou, Fleischmann, and Wallace}{Zhou
  et~al\mbox{.}}{2010}]%
        {Zhou2010}
\bibfield{author}{\bibinfo{person}{Yingjie Zhou}, \bibinfo{person}{Kenneth~R.
  Fleischmann}, {and} \bibinfo{person}{William~A. Wallace}.}
  \bibinfo{year}{2010}\natexlab{}.
\newblock \showarticletitle{{Automatic text analysis of values in the enron
  email dataset: Clustering a social network using the value patterns of
  actors}}. In \bibinfo{booktitle}{{\em Proceedings of the Annual Hawaii
  International Conference on System Sciences}}. \bibinfo{pages}{1--10}.
\newblock
\showISBNx{9780769538693}
\showISSN{15301605}
\showDOI{%
\url{https://doi.org/10.1109/HICSS.2010.77}}


\bibitem[\protect\citeauthoryear{Zhou, Goldberg, Magdon-Ismail, and
  Wallace}{Zhou et~al\mbox{.}}{2007}]%
        {Zhou2007}
\bibfield{author}{\bibinfo{person}{Yingjie Zhou}, \bibinfo{person}{Mark
  Goldberg}, \bibinfo{person}{M. Magdon-Ismail}, {and} \bibinfo{person}{W.a.
  Wallace}.} \bibinfo{year}{2007}\natexlab{}.
\newblock \showarticletitle{{Strategies for cleaning organizational emails with
  an application to enron email dataset}}. In \bibinfo{booktitle}{{\em 5th
  Conf. of North American Association for Computational Social and
  Organizational Science}}. \bibinfo{address}{Emory - Atlanta, Georgia}.
\newblock
\showISBNx{1518276822}
\showURL{%
\url{http://citeseerx.ist.psu.edu/viewdoc/download?doi=10.1.1.64.2123}}


\end{thebibliography}
\end{document}